\documentclass[journal=nanoletters,manuscript=article]{achemso}

\usepackage[version=3]{mhchem} 
\usepackage[T1]{fontenc}
\usepackage[colorlinks=true,citecolor=blue,linkcolor=blue,urlcolor=blue,anchorcolor=blue]{hyperref}%
\hypersetup{colorlinks=true,citecolor=blue,linkcolor=blue,urlcolor=blue}
\providecommand{\U}[1]{\protect\rule{.1in}{.1in}}
\usepackage{amsmath}
\usepackage{amssymb}
\usepackage{graphicx}
\usepackage{mathrsfs}
\usepackage{amsfonts}
\usepackage{amsthm}
\usepackage{color}
\usepackage{txfonts}

\author{Lingling Song}
\author{Huanhuan Yang}
\author{Yunshan Cao}
\author{Peng Yan}
\email{yan@uestc.edu.cn}
\affiliation[University of
Electronic Science and Technology of China]
{School of Electronic Science and Engineering and State Key Laboratory of Electronic Thin Films and Integrated Devices, University of Electronic Science and Technology of China, Chengdu 610054, China}

\title[An \textsf{achemso} demo]
  {Realization of the square-root higher-order topological insulator in electric circuits}

\abbreviations{IR,NMR,UV}

\begin{document}
\setlength{\baselineskip}{16pt}
\captionsetup[figure]{labelfont={bf},name={Figure},labelsep=period}
\textbf{Abstract:} Higher-order topological insulator (HOTI) represents a new phase of matter, the characterization of which goes beyond the conventional bulk-boundary correspondence and is attracting significant attention by the broad community. Using a square-root operation, it has been suggested that a square-root HOTI may emerge in a hybrid honeycomb-kagome lattice. Here, we report the first experimental realization of the square-root HOTI in topological LC circuits. We show theoretically and experimentally that the square-root HOTI inherits the feature of wave function from its parent, with corner states pinned to non-zero energies. The topological feature is fully characterized by the bulk polarization. To directly measure the finite-energy corner modes, we introduce extra grounded inductors to each node. Our results experimentally substantiate the emerging square-root HOTI and pave the way to realizing exotic topological phases that are challenging to observe in condensed matter physics.



\section{Introduction}
Topological insulators manifest unique and often counterintuitive properties, such as the robust electron transport and wave propagation against defects and disorder \cite{Hasan2010,Qi2011,Bernevig2013}, which are appealing for quantum computing interface \cite{Barik2018,Kitaev2001} and nonreciprocal lasing \cite{Bahari2017,Bandres2018,Harari2018,St-Jean2017}. Conventional topological insulators support chiral edge states of codimension one. In contrast, the emerging higher-order topological insulator (HOTI) allows topologically non-trivial boundary states with codimension larger than one \cite{Benalcazar2017,Benalcazar20172,Song2017,Ezawa2018}, thus shifting the standard paradigm of bulk-boundary correspondence. The realization of HOTIs has been reported in variety of fields, such as photonics \cite{Xie2018,Noh2018,Hassan2019,Mittal2019,Chen2019,Xie2019,Ota2019,ZhangL2019,LiM2019}, acoustics \cite{Xue2019,Ni2019,Xue2019_2,He2019,ZhangX2019,ChenZ2019,ZhangZ2019,Zhang2019}, mechanics \cite{Serra-Garcia2018,Fan2019}, and recently spintronics \cite{Li2019_1,Li2019_2,Li2019_3}. Using a square-root operation, Arkinstall \emph{et al.} proposed a pathway to design a new type of topological material, the so-called square-root topological insulator, which inherits the nontrivial nature of Bloch wave function from its parent Hamiltonian \cite{Arkinstall2017}. The approach is analogous to the case that Dirac discovered the positron when he took the square root of the Klein-Gordon equation for relativistic particles \cite{Dirac1928}. Photonic experiments indeed confirm the square-root topological insulator \cite{Kremer2020}. Very recently, Mizoguchi \emph{et al.} generalized the idea to HOTI by showing that the square-root HOTI may emerge in a decorated honeycomb lattice \cite{Mizoguchi2020}. However, the experimental realization of squre-root HOTI is still missing.

It has been shown that electrical circuits can act as a powerful platform to realize topological states \cite{Lee2018,Imhof2018,Garcia2019,Bao2019,Yang2020,Ezawa2018PRB} that are challenging to observe in condensed matter experiments. Here, we demonstrate the square-root HOTI in two-dimensional LC circuits arranged in a hybrid honeycomb-kagome (HK) manner [see Figure \ref{model_inf}a]. We show theoretically and experimentally that the square-root HOTI inherits features from its parent, with the topology being fully characterized by the bulk polarization. We find that the emerging corner states are pinned to finite energies, so that they cannot be directly observed by impedance measurements, in contrast to their ``zero-energy" counterpart in normal HOTIs \cite{Imhof2018,Yang2020}. To overcome this issue, we introduce extra grounded inductors, which shift the ``non zero-energy" corner modes to zero energy without modifying their wave functions.

\section{Results and discussion}

\subsection{Honeycomb-kagome electrical circuit}
We consider a lossless linear electric circuit consisting of capacitors and inductors and label the nodes of the circuit using $a=1,2,\cdots$. The responses of the circuit at frequency $\omega$ obey Kirchhoff's law
\begin{equation}
I_a(\omega)=\sum_bJ_{ab}(\omega)V_b(\omega),
\end{equation}
where $I_a$ is the external current flowing into node $a$, $V_b$ is the voltage of node $b$, and $J_{ab}$($\omega$) is the circuit Laplace operator
\begin{equation}
J_{ab}(\omega)=i\mathcal {H}_{ab}(\omega)=i\omega\left[C_{ab}+\delta_{ab}\left(\sum_nC_{an}-\frac{1}{\omega^2L_a}\right)\right],
\end{equation}
with $C_{ab} $ the capacitance between nodes $a$ and $b$ and $L_a$ being the grounding inductance of node $a$.
\begin{figure*}[htbp!]
 \centering

  \includegraphics[width=0.95\textwidth]{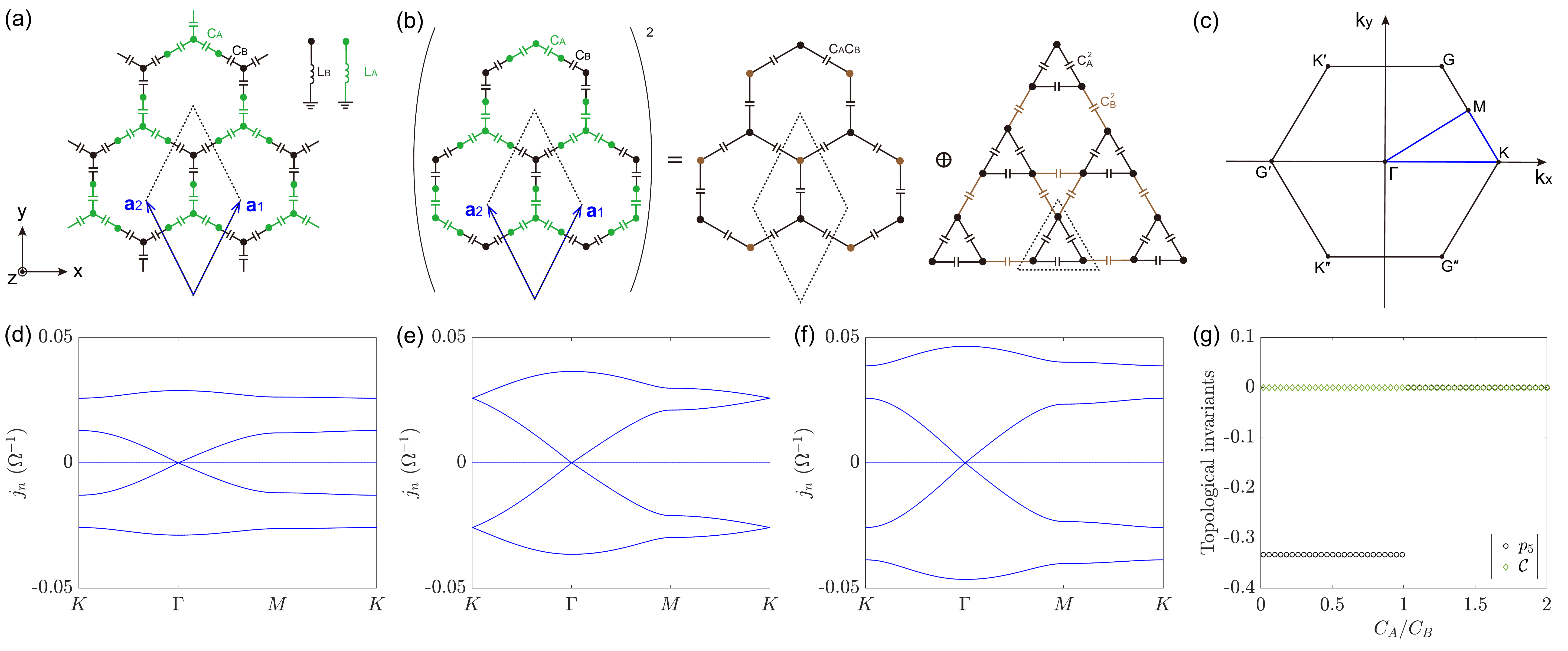}\\
  \caption{(a) Illustration of an infinite HK LC circuit. (b) The equivalence between the squared Hamiltonian of the HK circuit and its parent. (c) The first Brillouin zone. Band structures for different capacitance ratios: $C_A/C_B<1$ (d), $C_A/C_B=1$ (e), and $C_A/C_B>1$ (f). (g) Topological invariants as a function of $C_A/C_B$. Black circles and green diamonds represent the polarization and Chern number of the 5th band, respectively.}\label{model_inf}
\end{figure*}

Figure \ref {model_inf}a shows the structure of an infinite HK circuit, the squared Hamiltonian of which can be viewed as the direct sum of the Hamiltonians of a honeycomb and a breathing-kagome circuits (see analysis below), as illustrated in Figure \ref{model_inf}b. The unit cell including five nodes is represented by the dashed black rhombus. $ \textbf {a}_{1}=\frac {1} {2} d \hat {x} + \frac {\sqrt {3}} {2} d \hat {y} $ and $ \textbf { a}_{2}=-\frac {1} {2} d \hat {x} + \frac {\sqrt {3}} {2} d \hat {y} $ are the two basis vectors with $d$ being the lattice constant.  Figure \ref {model_inf}c displays the first Brillouin zone (BZ). The Hamiltonian of the circuit can be written as
\begin{equation}\label{Eq3}
 \mathcal {H}=\left(
 \begin{matrix}
   Q_{11} & 0        & Q_{1}& Q_{1} & Q_{1}\\
     0    & Q_{22}    & Q_{2} & Q_{3}^{*}& Q_{4}^{*}\\
   Q_{1}  & Q_{2}    & Q_{33} & 0 & 0 \\
   Q_{1}  & Q_{3}    &  0   & Q_{44}& 0 \\
   Q_{1}  & Q_{4}    &  0  & 0 & Q_{55}\\
  \end{matrix}
  \right),
\end{equation}
with matrix elements
\begin{equation}\label{Eq4}
\begin{aligned}
&Q_{11}=\omega\left(3C_{A}-\frac{1}{\omega^{2}L_A}\right),~Q_{22}=\omega\left(3C_{B}-\frac{1}{\omega^{2}L_B}\right),\\
&Q_{33}=Q_{44}=Q_{55}=\omega\left(C_{A}+C_{B}-\frac{1}{\omega^{2}L_A}\right), \\
&Q_{1}=-\omega C_{A},~Q_{2}=-\omega C_{B}, \\
&Q_{3}=-\omega C_{B}e^{-i\bf{k}\cdot{a}_1},~Q_{4}=-\omega C_{B}e^{-i\bf{k}\cdot{a}_2},\\
\end{aligned}
\end{equation}
with $\mathbf{k}=(k_x,k_y)$ the wave vector. By properly selecting $C_{A}$, $C_{B}$, $L_{A}$, and $L_{B}$, i.e., $C_{A}:C_{B}=L_{B}:L_{A}=1:2$, the diagonal elements of the Hamiltonian vanish at resonance. The Hamiltonian then can be simplified to
\begin{equation}\label{Eq5}
 \mathcal {H}=\left(
 \begin{matrix}
     O_{2,2} & \Phi_{\bf k}^{*}\\
    \Phi_{\bf k} & O_{3,3}\\
 \end{matrix}
 \right),
\end{equation}
where $O_{2,2}$ and $O_{3,3}$ are the $2\times2$ and $3\times3$ zero matrix, respectively, and $\Phi_{\bf k}$ is the $3\times2$  matrix
\begin{equation}\label{Eq6}
  \Phi_{\bf k}=\left(
 \begin{matrix}
   Q_{1}  & Q_{2} \\
   Q_{1}  & Q_{3} \\
   Q_{1}  & Q_{4}\\
 \end{matrix}
 \right).
\end{equation}

The Hamiltonian \eqref{Eq5} is chiral-symmetric, because $\mathcal {H}$ meets the condition $\mathcal {H} = -\gamma \mathcal{H}\gamma$ with
\begin{equation}\label{Eq7}
  \gamma=\left(
 \begin{matrix}
     I_{2,2} & O_{2,3}\\
    O_{3,2} & -I_{3,3}\\
 \end{matrix}
 \right),
\end{equation}
where $I_{2,2}$ and $I_{3,3}$ represent the $2\times2$ and $3\times3$ identity matrix, respectively. This indicates the existence of the parent Hamiltonian whose square root gives $\mathcal {H}$ \cite{Arkinstall2017,Mizoguchi2020}. Indeed, one can obtain the eigenvalues of $\mathcal {H}$ by taking its square
\begin{equation}\label{Eq10}
[\mathcal {H}]^{2}=\left(
 \begin{matrix}
     h_{\bf k}^{H} & O_{2,3}\\
    O_{3,2} & h_{\bf k}^{K}\\
 \end{matrix}
 \right),
\end{equation}
where $h_{\bf k}^{H}=\Phi_{\bf k}^{\dag}\Phi_{\bf k}$ and $h_{\bf k}^{K}=\Phi_{\bf k}\Phi_{\bf k}^{\dag}$ represent the Hamiltonian of a honeycomb sublattice with staggered on-site potentials and a breathing kagome sublattice, respectively, as plotted in Figure \ref{model_inf}b. The dispersion relation then can be conveniently computed by solving \eqref{Eq10}: $\varepsilon_{\bf k}^{H}=E_{\bf k}^{\pm}=3(C_{A}^{2}+C_{B}^{2}\pm
 \sqrt{4C_{A}^{2}C_{B}^{2}|\Delta({\bf k})|^{2}})/2$ for $h_{\bf k}^{H}$ and $\varepsilon_{\bf k}^{K}=0,E_{\bf k}^{\pm}$ for $h_{\bf k}^{K}$ with $\Delta({\bf k})=(1 + e^{i\bf{k}\cdot\bf{a}_1} + e^{i\bf{k}\cdot\bf{a}_2})/3 $, and taking a square-root operation afterwards. The band structures $\varepsilon_{\bf k}=0,\pm\sqrt{E_{\bf k}^{\pm}}$ for different capacitance ratios $C_{A}/C_{B}$ are shown in Figures \ref{model_inf}d-\ref{model_inf}f, which is gapped at $K$ point for $C_{A}/ C_{B}\neq 1$. It is therefore possible to realize the TI phase due to the gap opening. Interestingly, we find that the $\varepsilon_{\bf k}^{K}=0$ solution indicates a flat band. As shown in Figures \ref{model_inf}d-\ref{model_inf}f, the peculiar flat band touches other dispersive bands, which is an ideal system to study the noncontractible loop state and the topologically protected band-touching \cite{yan2020}.

The first-order topological insulator phase can be determined by computing the topologically invariant Chern number \cite{Avron1983,Wang2017}
\begin{equation}\label{Eq13}
   \mathcal{C}=\frac{i}{2\pi}\int\!\!\!\int_{\text{BZ}}dk_{x}dk_{y}\text{Tr}\left[P\left( \frac{\partial P}{\partial k_{x}}\frac{\partial P}{\partial k_{y}}- \frac{\partial P}{\partial k_{y}}\frac{\partial P}{\partial k_{x}}\right)\right],
\end{equation}
 where $P$ is the projection matrix $P(\textbf{k})=\phi (\textbf{k})\phi (\textbf{k})^{\dag}$, with $\phi (\textbf{k})$ being the normalized eigenstates of Eq. (\ref{Eq3}) in any band, and the integral is over the first BZ.

In order to judge the existence of HOTI, we can employ the bulk polarization as a topological invariant. For $C_3$ symmetric systems, the bulk polarization for the $n$th band can be written as \cite{Ni2019,Fang2012}
\begin{equation}\label{Eq14}
 2\pi p_{n}=\text{arg} \theta_{n}(\mathbf{k}=K)\ (\text{mod}~ 2\pi),
\end{equation}

where $\theta_{n}({\bf k})=u^{\dag}_{n}({\bf k})U_{{\bf k}}u_{n}({\bf k})$ with $u_{n}({\bf k})$ the $n$th eigenvector . The $U$-matrix is expressed as
\begin{equation}
U_{\bf k}=\left(
 \begin{matrix}
     1 & 0& 0& 0& 0\\
    0& e^{-i\bf{k}\cdot{a}_2}& 0& 0& 0\\
    0& 0& 0& 0& 1\\
    0& 0& 1& 0&  0\\
    0& 0& 0& 1&  0\\
 \end{matrix}
 \right).
 \end{equation}Here, we are interested in the 5th band. As shown in Figure \ref{model_inf}g, $p_5$ takes $-1/3$ for $C_A/C_B<1$, and 0 for $C_A/C_B>1$. A topological phase transition occurs at $C_A/C_B=1$. Meanwhile, the Chern number $\mathcal{C}$ vanishes for all $C_{A}/C_{B}$, indicating that the system does not support the first-order TI phase. We thus conclude that the system allows only two topologically distinct phases for $C_{A}/C_{B}<1$ and $C_{A}/C_{B}>1$.

\subsection{Corner states}

\begin{figure*}
  \centering
  \includegraphics[width=0.95\textwidth]{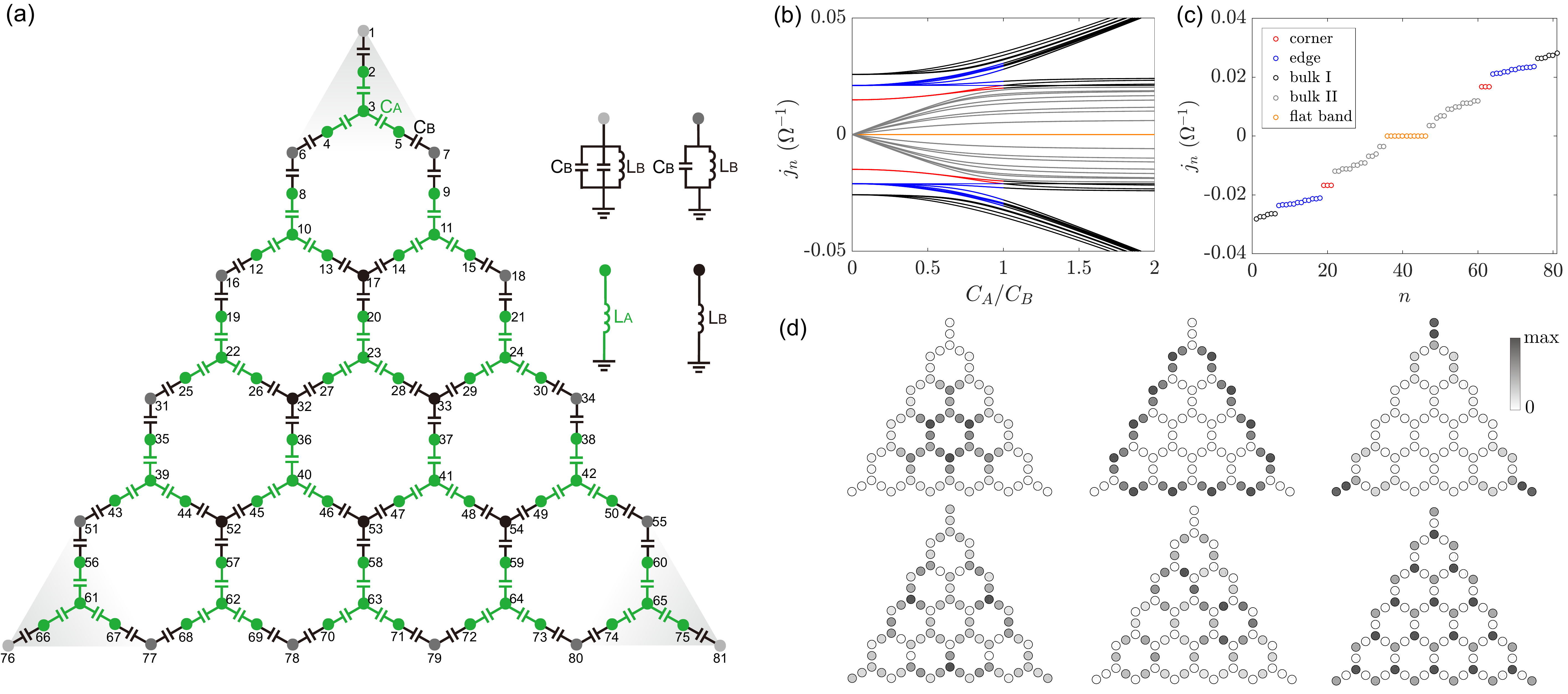}\\
  \caption{(a) Schematic plot of a finite-size HK circuit. (b) The admittance spectrum at the resonant condition under different ratios $C_A/C_B$ with the red segments denoting the corner-state phase. (c) Admittances for $C_A/C_B=0.5$. The red, blue, black, gray, and orange dots represent the corner, edge, bulk I, bulk II,  and flat-band  states, respectively. (d) Spatial distribution of the normalized bulk I, edge, corner, bulk II, and two flat-band modes (from left to right and top to bottom) with $j_n=-0.0282$, $-0.02108$, $-0.01668$, $-0.01106$, $-1.132\times10^{-18}$,  and $5.172\times 10^{-19}$ $\Omega^{-1}$, respectively.}\label{Energy_spectrum}
\end{figure*}

We consider a finite-size HK circuit with $\mathcal{N} = 81$ nodes, as depicted in Figure \ref{Energy_spectrum}a. The circuit Laplacian $J(\omega)$ reads
\begin{equation}\label{Eq15}
J(\omega)=\left(
\begin{array}{ccccccc}
J_{0B}  & -J_B  & 0    &  0   &   0   &   0 & \ldots\\
-J_B & J_{0A}   & -J_A & 0    &   0   &   0 & \ldots \\
0 & -J_A  & J_{0A}  &  -J_A   &  -J_A &   0 & \ldots\\
0    &  0 & -J_A    &  J_{0A} &  0    & -J_B & \ldots\\
0    & 0     & -J_A & 0   &  J_{0A} &   0 & \ldots\\
0    & 0     & 0 &  -J_B   & 0  &    J_{0B}& \ldots\\
\vdots  & \vdots  & \vdots &  \vdots   &  \vdots   &  \vdots& \ddots \\
\end{array}
\right)_{\mathcal{N}\times \mathcal{N}},
\end{equation}
with $J_{0{A(B)}}=3i\omega C_{A(B)}+1/(i\omega L_{A(B)})$ and $J_{A(B)}=i\omega C_{A(B)}$.

By diagonalizing \eqref{Eq15}, we obtain both eigenvalues $j_n$ (admittances) and eigenfunctions $\psi_n$ with $n=1,2,...,\mathcal{N}$. In Figure \ref{Energy_spectrum}b, we display the admittance spectrum for $C_A/C_B$ ranging from 0 to 2. The HOTI phase only appears for $0<C_A/C_B<1$ (the red line segments). In Figure \ref{Energy_spectrum}c, we exhibit a representative example at $C_A/C_B=0.5$, where the red, blue, black, gray, and orange dots represent the corner, edge, bulk I, bulk II,  and flat-band  states, respectively. We set $C_A=C_B/2=5$ nF and $L_A=2L_B=30$ $\mu$H in the calculation if not stated otherwise. It can be clearly seen that two groups of three-fold degenerate modes (red dots) are in the band gap, indicating the corner states. The spatial distribution of wave functions are plotted in Figure \ref {Energy_spectrum}d, from which we can distinguish these modes. Bulk II is spatially more extended than bulk I. In addition, we observe two kinds of flat-band modes: one lies in the kagome sublattice and the other one localizes in the honeycomb sublattice, in contrast to the infinity HK lattice, where flat-band states only appear in the kagome sublattice \cite{yan2020}.
 \begin{figure}[!htbp]
  \centering
  \includegraphics[width=0.6\textwidth]{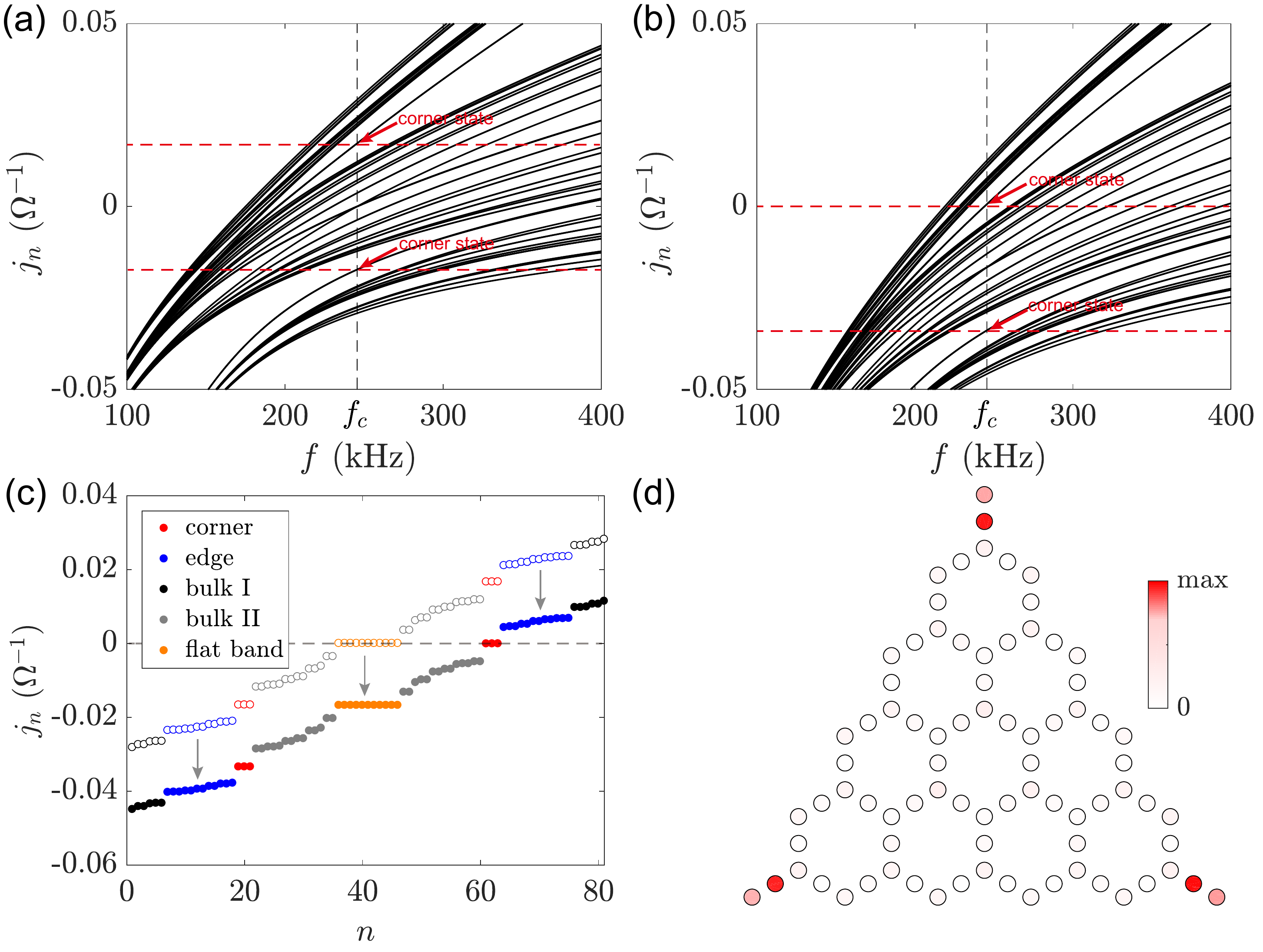}\\
  \caption{Theoretical spectrum of the circuit Laplacian $J(\omega)$ versus the driving frequency without (a) and with (b) grounded inductors  $L_G$. (c) Admittances for $C_A/C_B=0.5$ with $L_G=40$ $\mu$H. The red, blue, black, gray, and orange dots represent the corner, edge, bulk I, bulk II, and flat-band states, respectively. (d) The impedance distribution of corner state over the system.}\label{Energy_spectrum_f_Lg}
\end{figure}

Figure \ref{Energy_spectrum_f_Lg}a shows the theoretical spectrum of the circuit Laplacian $J(\omega)$ as a function of the driving frequency, where two isolated bands localize in the gap of the spectrum, representing the corner states. We note that the corner states in our system are not at the ``zero energy" as the ones in the normal HOTI circuit \cite{Yang2020}, as plotted in Figures \ref{Energy_spectrum}c and \ref{Energy_spectrum_f_Lg}a. One therefore cannot detect the corner states through a direct two-point impedance measurement. To solve this problem, we connect grounded inductors $L_G$ to all nodes to shift the whole spectrum down and move the corner modes to the zero energy, as shown in Figure \ref{Energy_spectrum_f_Lg}b, whereas the relative positions of the corner, edge, bulk and flat-band states do not change. Diagonalizing \eqref{Eq15} with $L_G$, i.e., adding $1/(i\omega L_G)$ to each diagonal element of matrix $J(\omega)$, we obtain the modified eigenvalues $j_n$ (admittances) and find that the corner states indeed emerge at the zero energy (the solid red dots), as plotted in Figure \ref{Energy_spectrum_f_Lg}c. Figure \ref{Energy_spectrum_f_Lg}d shows the spatial distribution of the impedance for the corner state. In the theoretical calculations, we adopt $L_G=40~\mu$H.

\subsection{Experimental observation}

\begin{figure}[htbp!]
  \centering
  \includegraphics[width=0.6\textwidth]{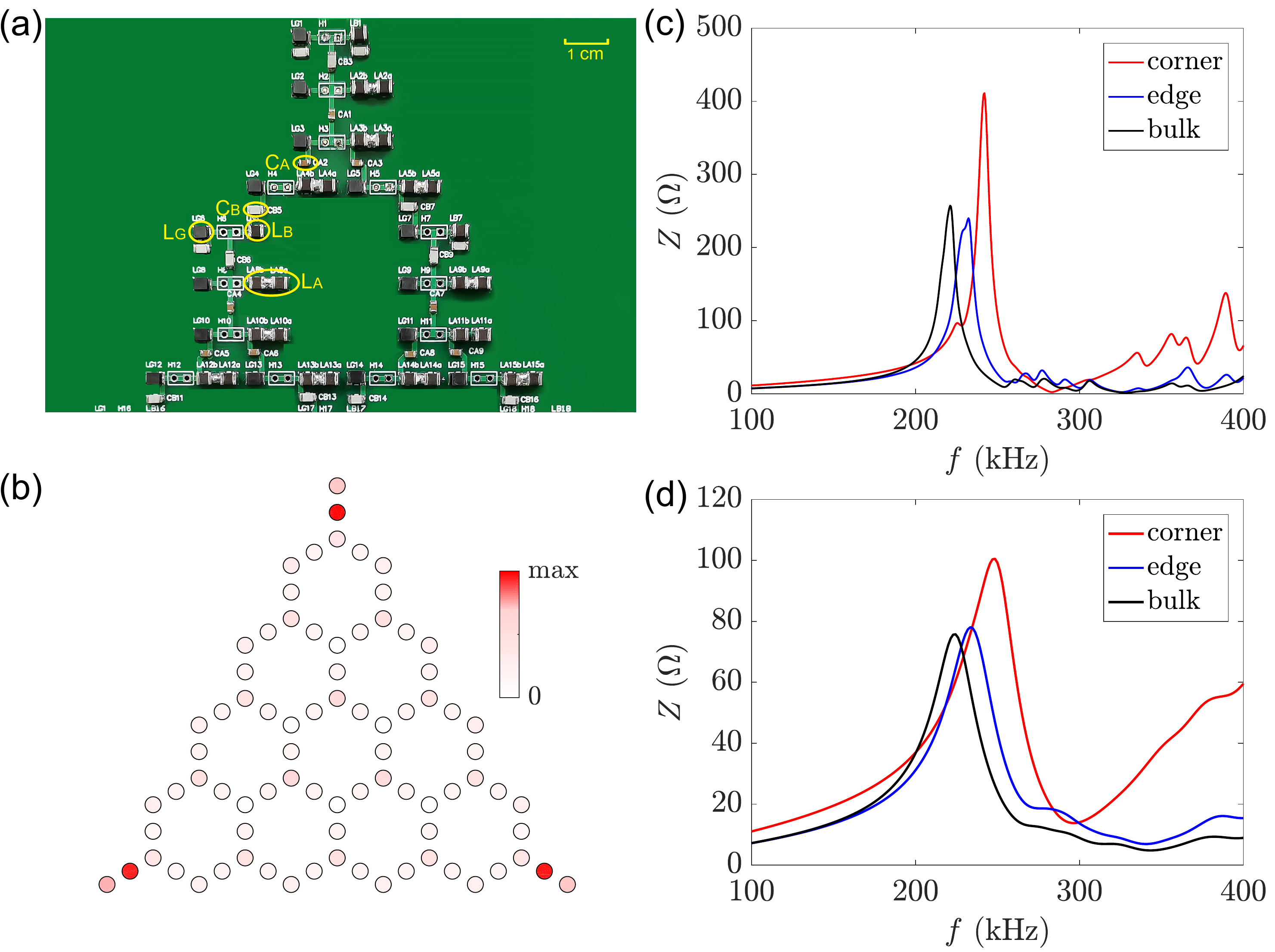}\\
  \caption{(a) Photograph of the partial layout of the experiment. (b) The distribution of the measured impedance over the system. (c) Theoretical impedance as a function of the driving frequency in a disordered circuit. (d) Measured impedance.}\label{Experiment}
\end{figure}

The electric circuits are fabricated on a printed circuit board as shown in Figure \ref{Experiment}a. We choose electric elements $C_A=C_B/2=5$ nF, $L_A=2L_B=28$ $\mu$H, and $L_G=39$ $\mu$H. The resonant frequency is given by $f_c=1/(2\pi\sqrt{3C_AL_A})=246$ kHz. It is noted that the inductance of $L_B$ decays slightly with the increasing of the driving frequency, with the mean value 14 $\mu$H near $f_c$. First of all, we measure the distribution of impedance in this circuit by the Impedance Analyzer (Keysight E4990A) [see Figure \ref{Experiment}b], which is in good agreement with the theoretical result in Figure \ref{Energy_spectrum_f_Lg}d. We then determine the impedance between three representative nodes and the ground as a function of the driving frequency, plotted in Figure \ref{Experiment}c (theory) and Figure \ref{Experiment}d (experiment), where we choose 2nd, 33th, and 34th nodes to quantify the signals from the corner, bulk, and edge states, respectively. In the theoretical calculation, we have averaged the result after ${10^3}$ realizations of uniformly distributed disorder (we assume a $2\%$ tolerance of each electric element). Both the theoretical and experimental (red) curves display a strong peak at the resonant frequency $f_c$, which confirms the very existence of the corner state.

\begin{figure}[htbp!]
  \centering
  \includegraphics[width=0.6\textwidth]{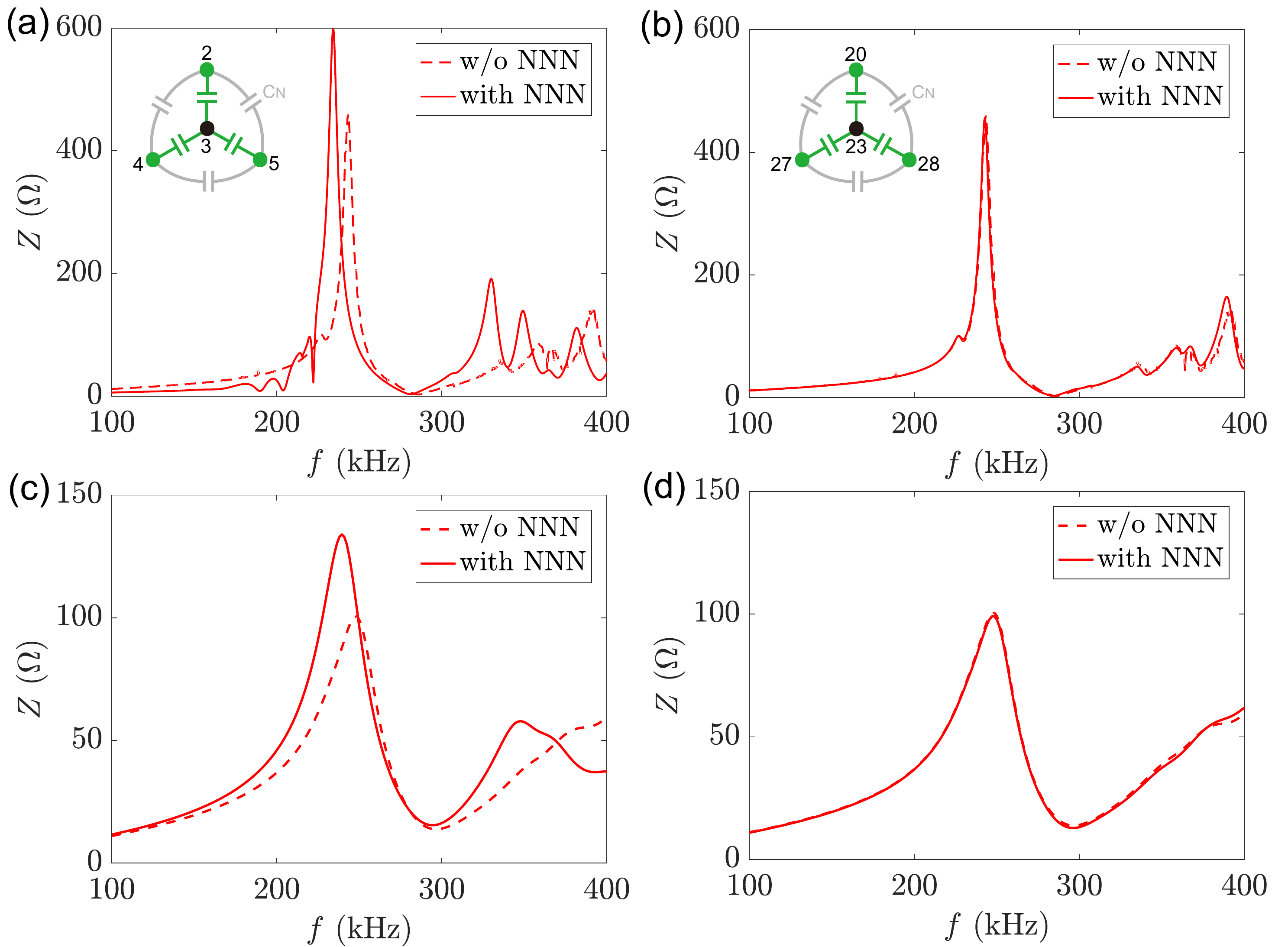}\\
  \caption{Theoretical impedance varying with the driving frequency with NNN hopping in the top corner (a) and in the deep bulk (b). Insets: configurations of the two types of NNN hopping. (c) and (d) Experimental results for the configurations shown in the insets of (a) and (b), respectively.}\label{nnn}
\end{figure}

To examine the robustness of the corner states, we introduce two kinds of next-nearest-neighbor (NNN) hopping terms to the system by connecting extra capacitors $C_N=2.2$ nF with the configurations shown in the insets of Figures \ref{nnn}a and \ref{nnn}b. When we connect 2nd, 4th, and 5th nodes [see Figure \ref{Energy_spectrum}a for numbered nodes] that breaks the chiral symmetry, the peak of zero mode suffers a blueshift about $9.3$ kHz, as depicted in Figures \ref{nnn}a and \ref{nnn}c. However, when we introduce NNN hopping in the deep bulk, by connecting 20th, 27th, and 28th nodes, the corner state remains at the same frequency because of the chiral symmetry conserving, as shown in Figures \ref{nnn}b and \ref{nnn}d. Experimental results agree well with theoretical calculations.

\section{Conclusion}
To summarize, we presented an experimental realization of the square-root HOTI in LC circuits. We demonstrated theoretically and experimentally that the square-root HOTIs inherit the feature of wave function from their parent. The emerging corner states are pinned to non-zero energies with topological features being fully characterized by the bulk polarization. To directly measure the finite-energy corner modes, we introduce extra grounded inductors to each node, thus shifting the localized modes to zero energy without modifying their wave distributions. Our results substantiate the emerging square-root HOTI and pave the way to observing exotic topological phases that are challenging to implement in condensed matter system.

\section{Acknowledgement}
This work was supported by the National Natural Science Foundation of China (Grants No. 11604041 and 11704060), and the National Key Research Development Program under Contract No. 2016YFA0300801.

\section{Abbreviation}
HOTI, Higher-order topological insulator; LC, inductors and capacitors; HK, honeycomb-kagome; BZ, Brillouin zone;
NNN, next-nearest-neighbor.

\section{Author information}
\subsection{Corresponding Authors}

\textbf{Peng Yan} - School of Electronic Science and Engineering and State Key Laboratory of Electronic Thin Films and Integrated Devices, University of Electronic Science and Technology of China, Chengdu 610054 China;
 Email: yan@uestc.edu.cn

\subsection{Authors}
\begin{flushleft}
\textbf{Lingling Song} - School of Electronic Science and Engineering and State Key Laboratory of Electronic Thin Films and Integrated Devices, University of Electronic Science and Technology of China, Chengdu 610054, China

\textbf{Huanhuan Yang} - School of Electronic Science and Engineering and State Key Laboratory of Electronic Thin Films and Integrated Devices, University of Electronic Science and Technology of China, Chengdu 610054, China

\textbf{Yunshan Cao} - School of Electronic Science and Engineering and State Key Laboratory of Electronic Thin Films and Integrated Devices, University of Electronic Science and Technology of China, Chengdu 610054, China
\end{flushleft}

\subsection{Author Contributions}
Peng Yan and Lingling Song conceived the idea and contributed to the project design. Lingling Song developed the theory and wrote the maniscript. Huanhuan Yang designed the circuits and performed the measurements. All authors discussed the results and revised the manuscript.
\subsection{Notes}
The authors declare no competing financial interest.


\end{document}